\documentclass[prl,twocolumn,showpacs,superscriptaddress]{revtex4}
\usepackage{graphicx}
\usepackage[tbtags]{amsmath}

\newcommand{\ij}{i\kern -0.08em j}

\begin{document}

\title{Rabi spectroscopy for a two-level system}
\author{Ya. S. Greenberg}
\affiliation{Novosibirsk State Technical University, 20 K. Marx
Ave., 630092 Novosibirsk, Russia}

\author{E. Il'ichev}
\affiliation{Institute for Physical High Technology, P.O. Box 100239, D-07702 Jena, Germany}

\date{\today}

%SQUIDs

\begin{abstract}

We have analyzed the interaction of a dissipative two level
quantum system with high and low frequency excitation. The system
is continuously and simultaneously irradiated by these two waves.
If the frequency of the first signal is close to the level
separation and the second one is tuned to the Rabi frequency, it
is shown that the response of the system exhibits an undamped low
frequency oscillation. The method can be useful for low frequency
Rabi spectroscopy in various physical systems which are described
by a two level Hamiltonian, such as nuclei spins in NMR, double
well quantum dots, superconducting flux and charge qubits, etc. As
an example, the application of the method to the readout of a flux
qubit is considered.
\end{abstract}

\maketitle

It is well known that under resonant irradiation, a quantum two
level system (TLS) can undergo coherent (Rabi) oscillations. Such
oscillations occur if microwaves are in resonance with the spacing
between the energy levels. In this case the level occupation
probability will oscillate with a frequency proportional to the
amplitude of resonant field~\cite{Rabi}. The effect is widely used
in molecular beam spectroscopy ~\cite{beam}, and in quantum
optics~\cite{Raimond}.

Normally, such oscillations are damped out with a rate, which is
dependent on how strongly the system is coupled to the
environment. In recent years, progress in the investigations of
scalable solid state qubits has necessitated the development of
different methods of advanced qubit control. One of the proposals
in this direction was to maintain the Rabi oscillations for
arbitrary long times by the synchronization of the phase of Rabi
oscillations, with that of the ideal solid state detector with the
aid of a quantum feedback loop ~\cite{Korotkov3}.

In this work we propose another method for maintaining such
oscillations. We consider a TLS which is irradiated continuously
by two external sources. The first one with a frequency
$\omega_0$, which is close to the energy gap between the two
levels, excites the low frequency Rabi oscillations. A second, low
frequency source tuned to the Rabi frequency causes the
oscillations to persist.

We start with a Hamiltonian of a TLS subjected to a both high and
low frequency excitation:
\begin{equation}\label{Ham1}
H = \frac{\Delta }{2}\sigma _x  + \frac{\varepsilon }{2}\sigma _z
- \sigma _z F\cos \omega _0 t - \sigma _z G(t).
\end{equation}
Hamiltonian ~(\ref{Ham1}) is written in the localized state basis.
Here  $\Delta$ is the tunnelling rate between localized states,
$\varepsilon$ is the bias, $\emph{F}$ is the coupling between the
TLS and high frequency field, and $\emph{G(t)}$ is a low frequency
external force. In the eigenstates basis, which we denote by upper
case subscripts for the Pauli matrices $\sigma_X, \sigma_Y,
\sigma_Z$, Hamiltonian ~(\ref{Ham1}) reads:
\begin{eqnarray}\label{Ham2}
H = \left[ {\frac{\Delta }{\Delta_{\varepsilon} }F\cos \omega _0 t
+
\frac{\Delta }{\Delta_{\varepsilon} }G(t)} \right]\sigma _X \nonumber\\
 +\left[{\frac{\Delta_{\varepsilon} }{2} - \frac{\varepsilon }{\Delta_{\varepsilon} }F\cos \omega _0 t
- \frac{\varepsilon }{\Delta_{\varepsilon} }G(t)} \right]\sigma
_Z,
\end{eqnarray}
where $\Delta_{\varepsilon} =\sqrt{\Delta^{2}+\varepsilon^{2}}$
is the gap between two energy states.

A more realistic description of the TLS requires the inclusion of
the dissipative environment in Hamiltonian (\ref{Ham1}), which for
a spin-boson model of the bath results in the Bloch-Redfield
equations for the matrix operators $\sigma_X$, $\sigma_Y$,
$\sigma_Z$, \cite{Bloch}, \cite{Hartmann}. For weak driving
($f\ll\Delta$), and for weak coupling of the TLS to the bath these
equations can be approximated by Bloch-type
equations~\cite{Saito}:

\begin{equation}\label{sigmaZ}
\left\langle {\dot \sigma _Z } \right\rangle  = \left( {2f\cos
\omega _0 t + 2g(t)} \right)\left\langle {\sigma _Y }
\right\rangle  - \Gamma _Z \left(\left\langle {\sigma _Z }
\right\rangle  - Z_0\right),
\end{equation}
\begin{align}\label{sigmaY}
\left\langle {\dot \sigma _Y } \right\rangle  =  - \left( {2f\cos
\omega _0 t + 2g(t)} \right)\left\langle {\sigma _Z }
\right\rangle \nonumber\\ + \left[ {\frac{\Delta_{\varepsilon}
}{\hbar } - \frac{{2\varepsilon }}{\Delta }f\cos \omega _0 t -
\frac{{2\varepsilon }}{\Delta }g(t)} \right]\left\langle {\sigma
_X } \right\rangle  - \Gamma \left\langle {\sigma _Y }
\right\rangle,
\end{align}
\begin{equation}\label{sigmaX}
    \left\langle {\dot \sigma _X } \right\rangle  =  - \left[ {\frac{\Delta_{\varepsilon} }{\hbar } -
    \frac{{2\varepsilon }}{\Delta }f\cos \omega _0 t - \frac{{2\varepsilon }}{\Delta }g(t)}
    \right]\left\langle {\sigma _Y } \right\rangle  - \Gamma \left\langle {\sigma _X }
    \right\rangle,
\end{equation}
where $f=\Delta$F$/\hbar\Delta_{\varepsilon}$,
$g(t)=\Delta$G(t)$/\hbar\Delta_{\varepsilon}$, and
$Z_0=-\tanh\left(\Delta_\varepsilon/k_BT\right)$ is the
equilibrium polarization of the system in the absence of external
excitation sources ($f=0$, $g=0$). The angled brackets in
Eqs.~(\ref{sigmaZ}), (\ref{sigmaY}), and (\ref{sigmaX}) denote the
trace over reduced density matrix $\rho(t)$, which is obtained by
tracing out all environment degrees of freedom:
$\langle\sigma_X\rangle=Tr(\sigma_X\rho(t))$, etc. In order to
simplify the problem we assume the  relaxation, $\Gamma_Z$ and
dephasing, $\Gamma$, rates in (\ref{sigmaZ})-(\ref{sigmaX}) are
time-independent, i.e. the rates are slowly varying functions on
the scale of Rabi period which is on the order of $1/f$.

 The desired solution of
Eqs.~(\ref{sigmaZ}), (\ref{sigmaY}), and (\ref{sigmaX}) is:
\begin{equation}\label{sZ}
    \left\langle {\sigma _Z } \right\rangle  = Z(t),
\end{equation}
\begin{equation}\label{sY}
    \left\langle {\sigma _Y } \right\rangle  = Y(t) + A(t)\cos (\omega _0 t) + B(t)\sin (\omega _0 t),
\end{equation}
\begin{equation}\label{sX}
    \left\langle {\sigma _X } \right\rangle  = X(t) + C(t)\cos (\omega _0 t) + D(t)\sin (\omega _0 t),
\end{equation}
where the quantities $X$, $Y$, $Z$, $A$, $B$, $C$, and $D$ are
slowly varying, compared to the high frequency $\omega_0$
operators. In the rotating wave approximation we obtain  from
Eqs.~(\ref{sigmaZ}), (\ref{sigmaY}), and (\ref{sigmaX}) the
response of the system to a small external force $g(t)$. The
Fourier components for the slowly varying $A(\omega )$, and
$B(\omega)$, and for the low frequency persistent response of the
TLS to a small low frequency excitation, $Z(\omega)$, $Y(\omega)$,
and $X(\omega)$ are as follows: $C(\omega)=B(\omega)$,
$D(\omega)=-A(\omega)$,

\begin{equation}\label{ZOmega}
Z(\omega ) = g(\omega )\frac{{2\varepsilon }}{\Delta }f^2 P_Z
Z_0\frac{\delta }{{\delta ^2  + \Gamma ^2 }}\frac{{2\Gamma  +
i\omega }}{s(\omega)},
\end{equation}
\begin{equation}\label{AOmega}
    A(\omega ) = g(\omega )\frac{{2\varepsilon }}{\Delta }f{\kern 1pt} P_Z Z_0\frac{\delta }{{\delta ^2  +
    \Gamma ^2 }}
     \frac{{\left( {2\Gamma  + i\omega } \right)\left( {i\omega  + \Gamma _Z }
     \right)}}{s(\omega)},
\end{equation}

\begin{widetext}
\begin{eqnarray}\label{BOmega}
 B(\omega ) =  - g(\omega )\frac{{2\varepsilon }}{\Delta }f{\kern 1pt} P_Z
 Z_0
 \frac{\delta }{{\delta ^2  + \Gamma ^2 }}
 \frac{1}{{\left( {\delta^2 +(\Gamma+i\omega)^2  } \right)}}
{\rm{}}\left\{ {\frac{{f^2 \delta \left( {2\Gamma + i\omega }
 \right)}}{s(\omega)} - \frac{{\delta ^2  - \Gamma ^2  - i\omega \Gamma }}{\delta }}
 \right\},
 \end{eqnarray}
\begin{equation}\label{YOmega}
Y(\omega ) = \frac{\hbar }{\Delta_{\varepsilon} }\frac{\varepsilon
}{\Delta }fA(\omega ) + \left( {\frac{\hbar }{\Delta_{\varepsilon}
}} \right)^2 \left( {2g(\omega )P_Z Z_0 - \frac{\varepsilon
}{\Delta }fB(\omega )} \right)\left( {\Gamma  + i\omega } \right),
\end{equation}
\begin{equation}\label{XOmega}
X(\omega ) = \left( {\frac{\hbar }{\Delta_{\varepsilon} }}
\right)^2 \frac{\varepsilon }{\Delta }fA(\omega )\left( {\Gamma  +
i\omega } \right) - \frac{\hbar }{\Delta_{\varepsilon} }\left(
{2g(\omega )P_Z Z_0 - \frac{\varepsilon }{\Delta }fB(\omega )}
\right),
\end{equation}
\end{widetext}
where  $\delta$ is a high frequency detuning,
$\delta=\omega_0-\Delta_{\varepsilon}/\hbar$, $\Omega_R$ is the
Rabi frequency, $\Omega_R=\sqrt{\delta^2+f^2}$, $s(\omega)=(\Omega
_R-\omega +i\Gamma)(\Omega _R+\omega -i\Gamma)(i\omega + \Gamma
_Z)
       - f^2 \left( {\Gamma _Z  - \Gamma }
      \right)$, and
\begin{equation}\label{Polrz}
    P_Z  = \frac{{\Gamma _Z \left( {\Gamma ^2  + \delta ^2 } \right)}}{{\Gamma _Z \left( {\Gamma ^2  +
    \delta ^2 } \right) + f^2 \Gamma }}.
\end{equation}
$P_ZZ_0$ is the nonequilibrium polarization, i. e., the steady
state population difference between the qubit energy levels in the
case when the high frequency excitation is applied to the qubit.
When deriving the expressions ~(\ref{ZOmega})- (\ref{XOmega}) we
assume the force $g$ to be small, thereby we neglect the higher
order terms of $g$. In addition, we keep only the terms which
oscillate within the bandwidth of the Rabi frequency and neglect
the terms which are of the order of $\hbar f/\Delta_\varepsilon$,
$\hbar \Gamma/\Delta_\varepsilon$, $\hbar
\Gamma_Z/\Delta_\varepsilon$.

As is seen from these expressions the persistent low frequency
oscillations of the spin components $Z(t)$, $X(t)$, $Y(t)$ appear
as the response to the low frequency external force, $g(t)$, only
in the presence of the high frequency excitation ($f\neq0$).

Exactly at resonance ($\delta=0$) $A(\omega)=0$, $Z(\omega)=0$,
but $B(\omega)\neq0$. Hence, as is seen from Eqs.~(\ref{sZ}),
(\ref{sY}), (\ref{sX}), in this case the population of the two
levels are equalized, spin circularly rotates in the $XY$ plane
with frequency $\omega_0$, with the center of the circle being
precessed with the Rabi frequency.

The key point of the method we described above is that it allows
for the detection of the high frequency response of a TLS at a
frequency which is much less than the gap frequency. The low
frequency dynamics of the quantities  $\left\langle {\sigma _X }
\right\rangle, \left\langle {\sigma _Y } \right\rangle$,
$\left\langle {\sigma _Z } \right\rangle$  bears information about
the relaxation, $\Gamma_Z$, and dephasing, $\Gamma$, rates. The
experimental realization of this method requires the measurement
of the difference of low frequency response of the system of
interest with and without high frequency excitation.

The detection scheme depends on the problem under investigation.
For example, the method
 can easily be adapted for NMR. As is seen from~(\ref{Ham1}), the NMR
 case corresponds to the polarization of the sample with the fields
$\triangle$ and $\varepsilon$ along the $x$ and $z$ axes,
respectively, with a high frequency excitation $Fcos(\omega_{0}t)$
and a low frequency probe $G(t)$ being applied along the $z$ axis.

Another example are the macroscopic quantun TLS, such as
superconducting qubits based on mesoscopic Josephson
junctions~\cite{Makhlin}. For these qubits the Rabi oscillations
have been detected through the statistics of switching
events~\cite{Nakamura, Vion, Martinis, Chiorescu}. In parallel,
the detection of coherent oscillations through a weak continuous
measurement which does not destroy the quantum coherence of the
TLS was proposed~\cite{Averin, Korotkov1, Korotkov2}. Generally,
the signature of such oscillations can be found in the noise
spectrum of the TLS, which was  recently demonstrated for a
superconducting flux qubit~\cite{Il'ichev}. Let us consider last
example in more detail, namely, the application of the method to a
persistent current qubit, which is a superconducting loop
interrupted by three Josephson junctions ~\cite{Mooij},
\cite{Orlando}. The basis qubit states have an opposing persistent
current and the operator of the persistent current in the qubit
loop reads $\widehat{I}_q=I_q\sigma_z$. In eigenstate basis the
average current, $\langle{\widehat{I}_q\rangle}$ is:
\begin{equation}\label{Curr}
\langle \widehat{I}_q\rangle  = \frac{{I_q }}{\Delta_{\varepsilon}
}\left( {\varepsilon \langle {\sigma _Z } \rangle  - \Delta
\langle {\sigma _X } \rangle } \right).
\end{equation}

This current can be detected through the variation of its magnetic
flux either by a DC SQUID ~\cite{Lupascu} or by a high quality
resonant tank circuit inductively coupled to the qubit
~\cite{Smirnov, Greenberg3, Grajcar}.

For the flux qubit the bias $\varepsilon$ is controlled by a dc
flux $\Phi_X$: $\varepsilon=E_Jf_X$, where
$f_X=(\Phi_X/\Phi_0-1/2)$, $E_J=\Phi_0I_q/2\pi$, and $\Phi_0=h/2e$
is the flux quantum. The qubit is inductively coupled to a high
quality resonant tank circuit with inductance $L_T$, capacitance
$C_T$, and quality factor $Q_T$. The mutual inductance between the
qubit and the tank $M=k(L_qL_T)^{1/2}$, where $k$ is the
dimensionless coupling parameter, and $L_q$ is the inductance of
the qubit loop. The tank is biased by a low (compared to the gap)
frequency current $I_b=I_0cos(\omega t)$. This readout circuit has
proven to be successful for the investigation of quantum
properties of the flux qubit ~\cite{Il'ichev, Greenberg3,
Grajcar}. The qubit+tank system is described by the Hamiltonian
$H+H_T$, where $H_T$ is the Hamiltonian of the tank, and $H$ is
given in ~(\ref{Ham1}), with $G(t)=MI_qI_T(t)$, where $I_q$ is the
current flowing in the qubit loop, and $I_T$ is the current in the
tank coil. The voltage across the tank, $V(t)=V_Tcos(\omega t+
\chi)$, is given by ~\cite{Smirnov}:
\begin{equation}\label{Volt}
    \ddot V + \gamma _T \dot V + \omega _T^2 V = M\omega _T^2 \frac{{d\langle\widehat{I}_q \rangle}}{{dt}} +
    \omega _T^2 L_T \dot I_b,
\end{equation}
where  $\gamma_T= \omega_T/Q_T$, and $\omega_T=(L_TC_T)^{-1/2}$
is the tank resonance frequency, which is tuned to the Rabi
frequency $(\omega_T\simeq\Omega_R)$.

 From Eqs.~(\ref{sigmaZ}), (\ref{sigmaY}), and
(\ref{sigmaX}) we obtain:
\begin{equation}\label{dtcurr}
  \frac{{d\langle\widehat{I}_q \rangle}}{dt} = \frac{I_q }{\Delta_\varepsilon }
  \left( { - \varepsilon \Gamma _Z \left[\left\langle
   \sigma _Z \right\rangle  - Z_0 \right]
 + \Delta \Gamma
   \left\langle {\sigma _X } \right\rangle  + \frac{\Delta_{\varepsilon} }{\hbar }
   \Delta \left\langle {\sigma _Y }
   \right\rangle } \right).
\end{equation}
In order to get the tank response we have to keep, in the right
hand side of ~(\ref{dtcurr}), only the low frequency parts of the
averaged Pauli operators:
\begin{widetext}
\begin{equation}\label{Vomega}
  V(\omega )\left( {\omega _T^2  - \omega ^2  + i\omega \gamma _T } \right) = \frac{M\omega _T^2 I_q}{\Delta_{\varepsilon}}
  \left( { - \varepsilon \Gamma _Z Z(\omega ) + \Delta \Gamma X(\omega ) + \frac{\Delta_{\varepsilon} }
  {\hbar }\Delta Y(\omega )} \right) + i\omega \omega _T^2 L_T
  I_0.
\end{equation}
By taking into account that $G(\omega)=MI_qI_T(\omega)$, where
$I_T(\omega)=-iV(\omega)/\omega L_T$, we obtain with the aid of
the expressions ~(\ref{ZOmega})- (\ref{XOmega}) the low frequency
detuning $\xi$, and the damping, $\Gamma_T$ of the tank:
\begin{equation}\label{ksi}
    \xi  = \omega _T^2  - \omega ^2  - 2\frac{{k^2 \omega _T^2 L_q I_q^2 \Delta ^2 }}{{\Delta_{\varepsilon} ^3 }}P_Z Z_0
    \left\{ {1 + \left( {\frac{\varepsilon }{\Delta }} \right)^2 f^2 \frac{\Delta_{\varepsilon} }{\hbar }\frac{\delta }
    {{\delta ^2  + \Gamma ^2 }}\frac{{f_1 (\omega )}}{{d(\omega )}}}
    \right\},
\end{equation}
\begin{equation}\label{GammaT}
    \Gamma _T  = \gamma _T  - 2\frac{{k^2 \omega _T^2 L_q I_q^2 \Delta ^2 }}{{\Delta_{\varepsilon} ^3 }}P_Z Z_0
    \left( {\frac{\varepsilon }{\Delta }} \right)^2 f^2 \frac{\Delta_{\varepsilon} }{\hbar }\frac{\delta }{{\delta ^2
     + \Gamma ^2 }}\frac{{f_2 (\omega )}}{{d(\omega )}}.
\end{equation}
The quantities $\xi$ and $\Gamma_T$ are related to the voltage
amplitude $V_T$, and the phase, $\chi$: $V_T = \omega \omega _T^2
L_T I_0 /\sqrt {\xi ^2  + \omega ^2 \Gamma _T^2 }$, and $\tan\chi
= \xi /\omega \Gamma _T$. The functions $f_1$, $f_2$, and $d$,
which account for the resonant properties of the Rabi
oscillations, are as follows:
\begin{equation}\label{f1}
 f_1 (\omega ) = \left[ {\left( {2\Gamma \Gamma _Z  + \omega ^2 } \right)\left( {\Omega_R^2+\Gamma^2-
 \omega ^2 } \right) + 2\omega ^2 \Gamma \left( {\Gamma _Z  - 2\Gamma } \right) - 2f^2 \Gamma
 \left( {\Gamma _Z  - \Gamma } \right)} \right],
\end{equation}
\begin{equation}\label{f2}
    f_2 (\omega ) = \left[ {\Gamma _Z \left( {\Omega _R^2+\Gamma^2 - \omega ^2 } \right) + f^2
    \left( {\Gamma  - \Gamma _Z } \right) - 2\Gamma \left( {\Omega _R^2+\Gamma^2+ 2\Gamma \Gamma _Z }
    \right)} \right],
\end{equation}
\begin{equation}\label{d}
    d(\omega ) = \left[ {\left( {\Omega _R^2+\Gamma^2 - \omega ^2 } \right)^2  + 4\omega ^2 \Gamma ^2 }
     \right]\left( {\omega ^2  + \Gamma _Z^2 } \right) + f^4 \left( {\Gamma _Z  - \Gamma }
      \right)^2  - 2f^2 \left( {\Gamma _Z  - \Gamma } \right)\left[ {\Gamma _Z
      \left( {\Omega _R^2+\Gamma^2 - \omega ^2 } \right) - 2\omega ^2 \Gamma }
      \right].
\end{equation}

\end{widetext}
 The equations ~(\ref{ksi}, \ref{GammaT})
are obtained for the case $\Delta\delta/\hbar\gg f^2$. In the
opposite case, in particular, for zero high frequency detuning
($\delta=0$) the result is:
\begin{equation}\label{ksi0}
    \xi  = \omega _T^2  - \omega ^2  - 2\frac{{k^2 \omega _T^2 L_q I_q^2 \Delta ^2 }}{{\Delta_{\varepsilon} ^3 }}P_0 Z_0
    \left\{1 + \left( {\frac{\varepsilon }{\Delta }} \right)^2  \frac{f^2}{\Gamma^2+\omega^2}\right\},
\end{equation}
\begin{equation}\label{GammaT0}
    \Gamma _T  = \gamma _T  + 2\frac{{k^2 \omega _T^2 L_q I_q^2 \Delta ^2 }}{{\Delta_{\varepsilon} ^3 }}P_0 Z_0
    \left( {\frac{\varepsilon }{\Delta }} \right)^2  \frac{f^2}{\Gamma(\Gamma^2+\omega^2)}.
\end{equation}
where $P_0=P_Z(\delta=0)=\Gamma_Z\Gamma/(\Gamma_Z\Gamma+f^2)$.

\begin{figure}
\centering \includegraphics[width=6cm]{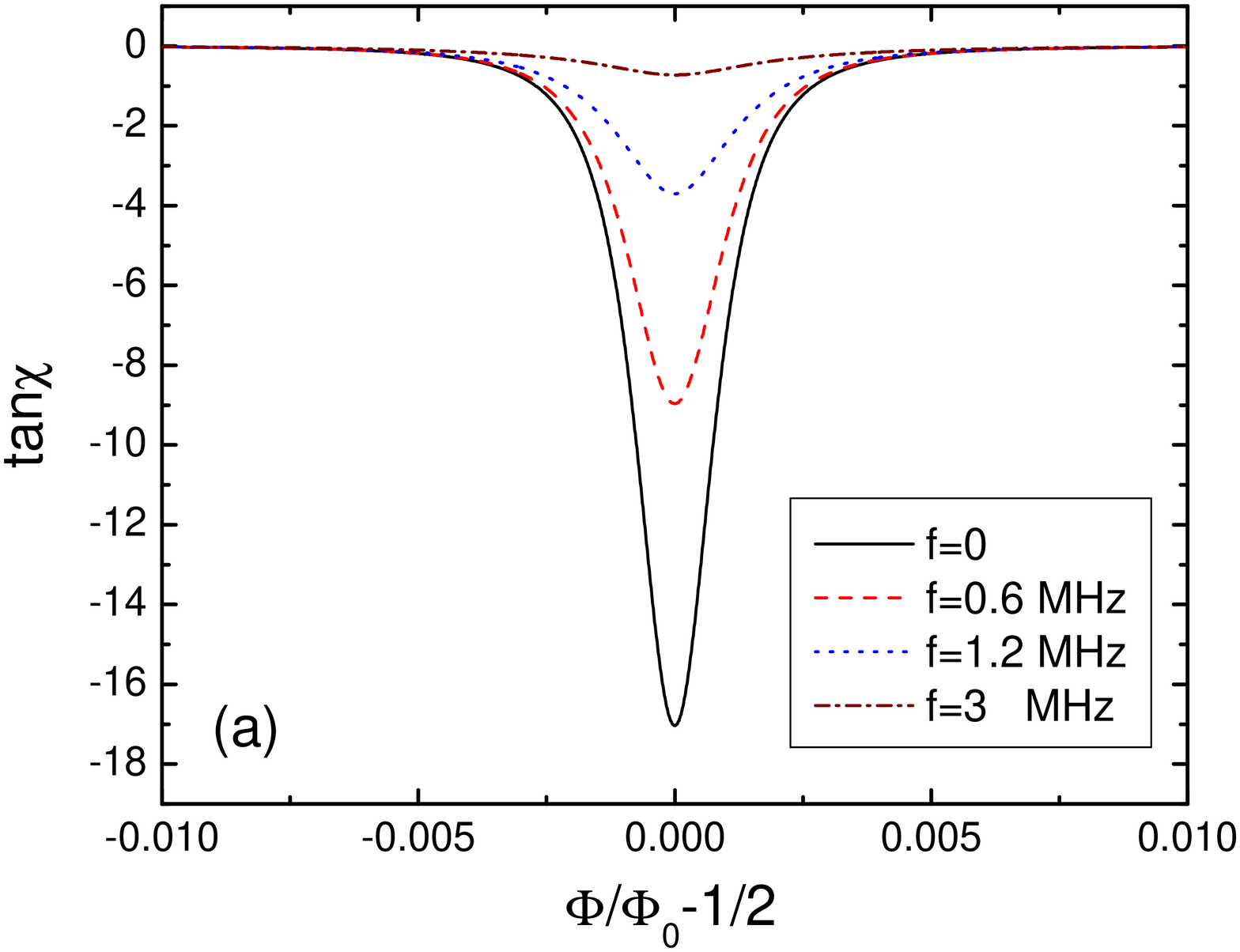} \centering
\includegraphics[width=6.cm]{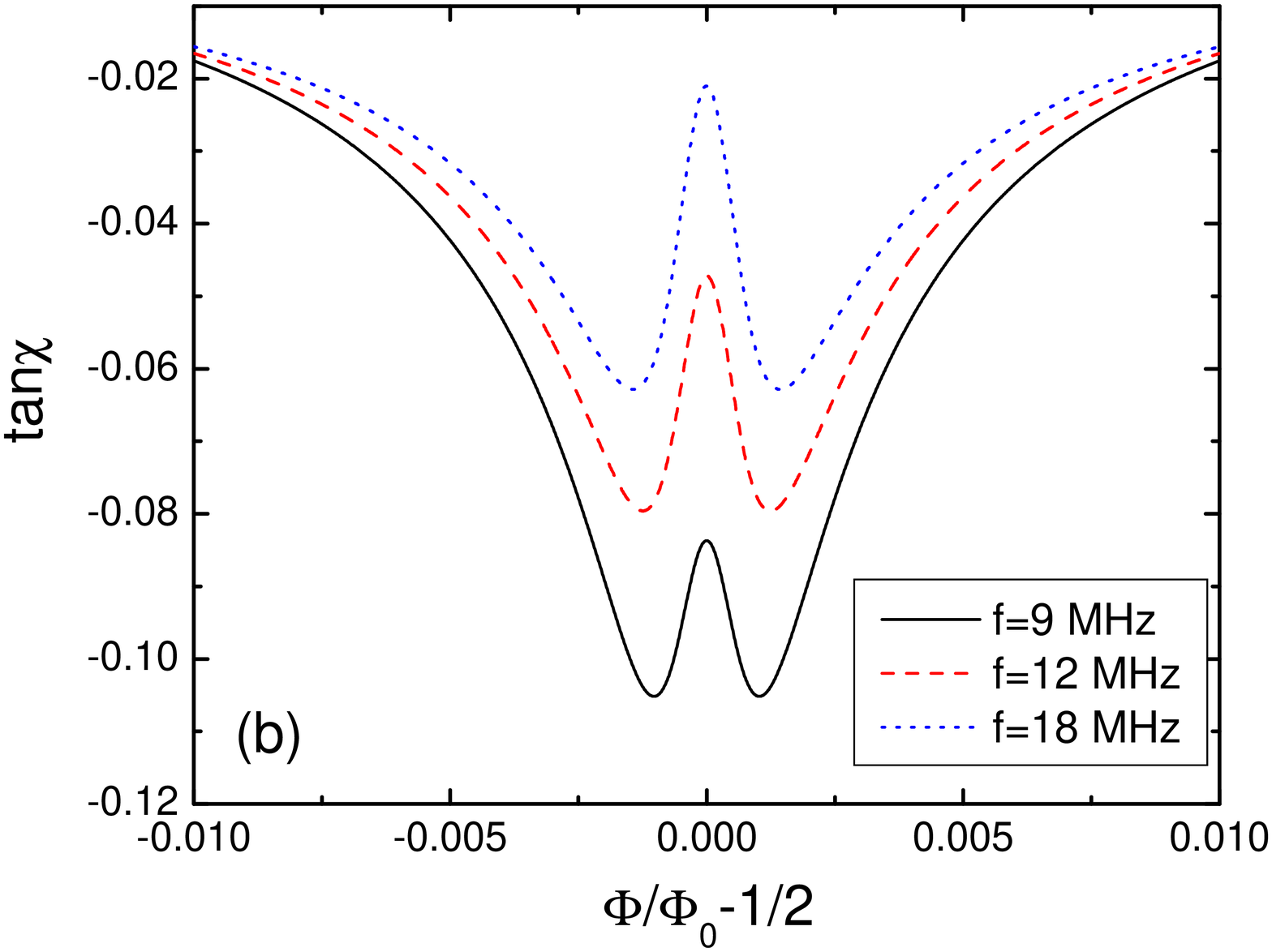}
\caption{The dependence of the tank phase on the flux when under
the influence of high frequency irradiation. $f=F/h$ is the power
of irradiation
 in frequency units.}
 \label{fig1}
\end{figure}

In the absence of high frequency excitation ($f=0)$ we obtain from
Eqs.~(\ref{ksi},~\ref{GammaT}) (or from
Eqs.~(\ref{ksi0},~\ref{GammaT0}) a purely inductive response
($\Gamma_T=\gamma_T$, $P_Z=1$), which describes the adiabatic
evolution of the qubit in the ground state ~\cite{Greenberg3}. The
experimental study of the adiabatic evolution provides us with
information about the energy gap between the two
levels~\cite{Grajcar}. Therefore, the difference of the low
frequency response of the persistent current qubit with and
without high frequency excitation can provide additional
information about the qubit's damping rates, $\Gamma$ and
$\Gamma_Z$ (see Eqs.~\ref{ksi}, \ref{GammaT}). As an illustration
of the method we show in Fig.~\ref{fig1} how the dependence of the
tank phase $\chi$ on the flux changes with the application of the
high frequency excitation. The graphs have been calculated for
zero high frequency detuning ($\delta=0$) from Eq.~(\ref{ksi0})
for the following parameters: $\omega/2\pi=\omega_T/2\pi=6$~MHz,
$k=0.03$, $Q_T=2000$; $E_J=1.32\times10^{-22}J$,
$E_C=2.14\times10^{-24}J$, $L_q=40$~pH, $I_q=280$~nA,
$\Delta/2\pi=1$~GHz, $\Gamma_Z/2\pi=0.1$~MHz, $\Gamma/2\pi=4$~MHz,
and $T=10$~mK. At a relatively low power of the irradiation the
form of the curves remains unchanged, with the amplitudes of the
dips being conditioned by the factor $P_0$ in Eq.~(\ref{ksi0})
(Fig.~\ref{fig1}a). At higher powers the second term in the curled
brackets of Eq.~(\ref{ksi0}), which describes the influence of
Rabi oscillations, comes into play. As a result, the forms of the
phase curves are changed drastically (Fig.~\ref{fig1}b).

In conclusion, we propose a method to maintain the Rabi
oscillations in a dissipative TLS by irradiating it with the two
external sources. The first, high frequency source, which is tuned
to the energy gap of the system, excites low frequency Rabi
oscillations. The second, low frequency source, which is tuned to
the frequency of the Rabi oscillations, causes the oscillations to
persist. When applied to a readout of the flux qubit inductively
coupled to a tank circuit, the method allows for the experimental
determination of the relaxation and dephasing rates of the qubit.

\begin{acknowledgments}
We are grateful to D. Averin, M. Grajcar, A. Korotkov, A.
Shnirman, and A. Smirnov  for a detailed discussion of the
problem. We thank also A. Izmalkov, W. Krech, A. Maassen van den
Brink, V. Shnyrkov,  Th. Wagner, and A. Zagoskin, for fruitful
discussions.

The authors acknowledge the support from D-Wave Systems. Ya.~G.
acknowledges partial support by the INTAS grant 2001-0809. E.~I.
thanks the EU for support through the RSFQubit project.

\end{acknowledgments}


\begin{thebibliography}{99}

\bibitem{Rabi} I. I. Rabi, Phys. Rev. \textbf{51}, 652 (1937).

\bibitem{beam} Atomic and Molecular Beams: The State of The Art
2000, (Roger Compargue, ed.) Springer Verlag Telos, 2001.

\bibitem{Raimond} J. M. Raimond, M. Brune, and S. Haroche, Rev. Mod. Phys. \textbf{73}, 565 (2001).



\bibitem{Korotkov3} A. N. Korotkov, cond-mat/0404696.
\bibitem{Bloch} F. Bloch, Phys. Rev.\textbf{105}, 1206 (1957); A. G. Redfield, IBM J. Res. Dev. \textbf{1}, 19 (1957).
\bibitem{Hartmann} L. Hartmann et al., Phys. Rev. E \textbf{61}, R4687 (2000).
\bibitem{Saito} S. Saito et al., cond-mat/0403425 (2004).
\bibitem{Makhlin} Y. Makhlin, G. Schon, and A. Shnirman, Rev. Mod. Phys. \textbf{73}, 357 (2001).

\bibitem{Nakamura} Y. Nakamura, Yu.A. Pashkin, and J.S. Tsai.,
Phys. Rev. Lett. \textbf{87}, 246601 (2001).

\bibitem{Vion} D. Vion et al., Science \textbf{296}, 886 (2002).

\bibitem{Martinis} J.M. Martinis et al, Phys. Rev. Lett. \textbf{89}, 117901 (2002).%S. Nam, J. Aumentado, and C. Urbina,

\bibitem{Chiorescu} I. Chiorescu et al.,Science \textbf{299}, 1869 (2003). %Y. Nakamura, C.J.P.M. Harmans, and J.E. Moo\ij,

\bibitem{Averin} D. V. Averin, cond-mat/0004364.

\bibitem{Korotkov1} A.N. Korotkov Phys. Rev. B \textbf{64}, 165310 (2001).

\bibitem{Korotkov2} A.N. Korotkov and D.V. Averin, Phys. Rev. B \textbf{63}, 115403 (2001).


\bibitem{Il'ichev} E. Il'ichev et al., Phys. Rev. Lett. \textbf{91}, 097906 (2003).

\bibitem{Mooij} J.E. Mooij  et al., Science \textbf{285}, 1036 (1999).
\bibitem{Orlando} T. P. Orlando et al., Phys. Rev. B \textbf{60}, 15398
(1999).
\bibitem{Lupascu} A. Lupascu, C. J. P. M. Harmans, and J. E. Mooij, cond-mat/0410730.

\bibitem{Smirnov} A. Yu. Smirnov, Phys. Rev. B \textbf{68}, 134514 (2003).

\bibitem{Greenberg3} Ya. S. Greenberg et al., Phys. Rev B \textbf{66}, 214525 (2002).


\bibitem{Grajcar} M. Grajcar et al., Phys. Rev. B \textbf{69}, 060501(R) (2004).



\end{thebibliography}
\end{document}